\documentclass[fleqn,10pt]{wlscirep} \title{Collective Influence of Multiple Spreaders Evaluated by Tracing Real Information Flow in Large-Scale Social Networks}
\author[1]{Xian Teng}
\author[2]{Sen Pei}
\author[1]{Flaviano Morone}
\author[1,*]{Hern{\'a}n A. Makse}
\affil[1]{Levich Institute and Physics Department, City College of New York, New York, NY 10031, USA}
\affil[2]{Department of Environmental Health Sciences, Mailman School of Public Health, Columbia University, New York, NY 10032, USA}

\affil[*]{hmakse@lev.ccny.cuny.edu}


\begin{abstract}

Identifying the most influential spreaders that maximize information
flow is a central question in network theory. Recently, a scalable method called ``Collective Influence (CI)'' has been put forward through collective influence maximization. In contrast to heuristic methods evaluating nodes' significance separately,
CI method inspects the collective influence of multiple
spreaders. Despite that CI applies to the influence maximization
problem in percolation model, it is still important to examine its
efficacy in realistic information spreading. Here, we examine
real-world information flow in various social and scientific platforms including
American Physical Society, Facebook, Twitter and LiveJournal. Since
empirical data cannot be directly mapped to ideal multi-source
spreading, we leverage the behavioral patterns of users extracted from
data to construct ``virtual'' information spreading processes. Our results demonstrate that the set of spreaders selected by
CI can induce larger scale of information
propagation. Moreover, local measures as the number of connections or
citations are not necessarily the deterministic factors of nodes'
importance in realistic information spreading. This result has
significance for rankings scientists in scientific networks like the APS, where
the commonly used number of citations can be a poor indicator of the
collective influence of authors in the community.

\end{abstract}
\begin{document}


\flushbottom
\maketitle
\thispagestyle{empty}

\section*{Introduction}

Identification of the most influential nodes in social networks has broad applications in a variety of network dynamics \cite{valente1999accelerating,domingos2002mining,vandenBulte2007new,iyengar2011opinion,watts2002asimple,watts2007influentials,albert2000error,pastor2001epidemic,newman2002spread,morone2016robust,yan2014dynamical,yan2015global}. For example, in viral marketing, advertising a small group of influential customers to adopt a new product can inexpensively trigger a large scale of further adoption \cite{valente1999accelerating,domingos2002mining,vandenBulte2007new,iyengar2011opinion}; in epidemics control, the immunization of structurally important persons can efficiently halt global epidemic outbreaks in contact networks \cite{pastor2001epidemic,newman2002spread,yan2014dynamical,yan2015global}; and in biological systems like brain networks, some significant nodes are responsible for broadcasting information and therefore locating and protecting them are crucial for the whole information processing system \cite{morone2016robust}. Given its practical significance, the problem of finding the optimal set of influencers in a given network has attracted much attention in network science \cite{leskovec2007cost,kempe2003maximizing,altarelli2013optimizing}.

For a long time, researchers have developed numerous heuristic measures as predictors of nodes' importance in information spreading. Among the most frequently used topological properties are the number of connections (degree) \cite{albert2000error,pastor2001epidemic}, betweenness \cite{freeman1977aset} and eigenvector centralities \cite{freeman1978centrality}, PageRank \cite{brin1998the}, k-core \cite{dorogovtsev2006kcore,carmi2007amodel,kitsak2010identification,pei2014searching,sen2013spreading}, etc. All of them are established in the non-interacting setting, where nodes' significance is evaluated by taking them as isolated agents. As a result, these ad-hoc approaches, designed for finding single superspreaders, fail to provide the optimal solution for the general case of multiple influencers. To address this many-body issue, a rigorous theoretical framework based on collective influence (CI) theory has recently been presented\cite{morone2015influence,pei2016collective}. With a broader notion of influence -- collective influence, the CI method pursues the goal of maximizing the overall influence of multiple spreaders. Such explicit optimization objective enables CI to give the minimal set of spreaders.

Although CI exhibits good performance with scalability in the optimal percolation model, more validation work regarding its efficacy in real-world information spreading still needs to be done. Previously, the lack of real data of information diffusion has led to the mainstream adoption of artificial spreading models to simulate spreading dynamics. However, the over-simplified spreading models usually neglect such important factors as activity frequency \cite{muchnik2013origins}, connection strength and behavioral preferences, thus fail to reproduce some observed characteristics of real information spreading \cite{goel2012the}. More importantly, different models may produce model-dependent contradictory results \cite{pei2014searching}. Therefore, it is necessary to evaluate CI's performance empirically through realistic information diffusion before applying it to real-world applications like marketing and advertising.

Here, we address this problem by tracking and analyzing the real-world
information flows in a wide range of social media: journals of
American Physical Society (APS), an online social network Facebook.com
(Facebook) \cite{viswanath2009on}, a microblogging service Twitter.com
(Twitter) and a blog website LiveJournal.com (LiveJournal). Rather
than tracking the spreading range of single spreaders
\cite{pei2014searching}, we intend to investigate the overall
spreading range, i.e., the collective influence, of multiple
spreaders. To achieve this, the most straightforward idea is to
extract and examine the real instances of information diffusion that
are triggered by multiple spreaders. Unfortunately, such ideal
multi-source spreading instances in which spreaders send out the same
piece of message at the same time rarely exist in reality. Even though
we can find such instances, the initial spreaders are hardly the same
as the set of nodes selected by CI or other heuristic strategies,
making the comparison between those methods impossible.

To overcome the aforementioned difficulties, we construct ``virtual''
multi-source spreading processes by following users' behavioral
patterns in the data. In particular, under the assumption that users
will maintain their personal preferences in spreading processes, we
measure the strength of directed social ties shown in historical
diffusion records to represent the influence strength of a user
imposing on another. For a node under influence of several spreaders,
the overall influence on it is defined as the highest influence
strength. In this way, we are able to quantify the collective
influence imposing on the entire network, corresponding to the
collective spreading range of virtual processes initiated by any given
set of seeds. Through comparisons with competing heuristic methods,
including high degree (HD) \cite{albert2000error,pastor2001epidemic},
adaptive high degree (HDA) \cite{morone2015influence}, PageRank (PR)
\cite{brin1998the} and k-core
\cite{dorogovtsev2006kcore,carmi2007amodel,kitsak2010identification,pei2014searching},
we find that the set of spreaders selected by CI can
exert larger collective influence on the population with the same
number of initial seeds. This provides a direct empirical validation
of CI's good performance in real information spreading. In addition,
some individual properties such as the number of connections and
citations, which were previously regarded as reliable predictors of
influence, are found to be invalid in the context of collective
influence. This in turn reflects that it is the interplay between
spreaders that determines the collective influence rather than
individual features.

\section*{Results}

\subsection*{Introduction of Datasets}

In the following empirical study, four datasets are examined: the
journals of American Physical Society (APS), an online social network
Facebook.com (Facebook) \cite{viswanath2009on}, a microblogging
service Twitter.com (Twitter), and a blog website LiveJournal.com
(LiveJournal). All datasets are available at
\url{kcore-analytics.com}. During the period of data collection,
people not only maintain social relations with their friends but also
interact with others to spread and receive information. Certainly,
there are diverse manifestations with respect to the social relation
and interaction in distinct platforms. For instance, in the academic
data of APS, authors show their social relations, i.e. coauthorship,
through jointly publishing articles, and they reveal their
interactions and information transmission by citing others'
papers. While in the online social media like Facebook, Twitter and
LiveJournal, users reflect their social relations by becoming ``cyber
friends'', and they interact with each other by creating, receiving,
and transmitting messages \cite{cheng2014can}. With the collection of
such information, we can obtain the full network structure as well as
the empirical information flows. Details about these data are
explained as follows.

\begin{itemize}
\item The American Physical Society (APS) is the world's largest organization of physicists. APS data contains the information of all the scientific papers published on APS journals until 2005, including Physical Review A, B, C, D, E and Physical Review Letters. From the author lists and references of scientific publications, we can obtain the information about collaborations and citations. In total, there are 299,996 articles and 230,521 authors in the data, along with 2,356,525 records of citations. We construct the underlying collaboration network according to their coauthorship. If two authors have published one article together, one undirected edge is built between them. Beyond that, we trace the information diffusion based on the reference flows. If a scientist $i$ cites one paper written by $j$, then we can say that information spreads from $j$ to $i$.

\item Facebook is an online social networking service. In Facebook, each registered user maintains a friend list, which is a good representation of actual social relationships. Users can exchange messages, post status updates and photos, share videos, and browse the posts published by their friends. The Facebook data contains the friend lists and the entire records of wall posts from the New Orleans regional network, over a period of two years from September 26th, 2006 to January 22nd, 2009. This data contains 63,731 users and 838,092 wall posts in total. The social network is extracted from the friend lists. If user $j$ is added into user $i$'s friend list (or $i$ is in $j$'s friend list), we assume that they are friends so that we build an undirected edge between them. According to the wall posts, we can infer the information diffusion flows. If user $i$ makes comments on user $j$'s page, we presume that $i$ has gained information from $j$ to motivate him/her to write comments.

\item Twitter is a microblogging service that enables users to send and read short word-limited messages called ``Tweets''. In the 2016 election year, Donald Trump, who is the presumptive nominee of the Republican Party for President of the United States, has become one of the most popular topics being discussed in Twitter. From February 10, 2016 to March 14, 2016, we collect approximately 670,000 Tweets that contain the key word ``Donald Trump'' or ``Trump''. In the collection of Tweets, we extract four kinds of Tweets: mention, replies, retweet and quote. A mention is a Tweet that contains another user's @username anywhere in the body of the Tweet. A reply is a response to another user's Tweet that begins with the @username of the person you're replying to. Replies are also considered as mentions. Besides, a retweet is a re-posting of someone else's Tweet, in which such character RT@username appears at the beginning to indicate that users are re-posting others' content. A quote is a special form of retweet that users can write their own comments when they are re-posting. We consider the mention (and also reply) relationship as a representative of strong social ties and use them to construct the network structure. Meanwhile, we use retweets (and quotes) to obtain information flows. If user $i$ retweets a Tweet from user $j$, we assume information diffuses from user $j$ to user $i$.

\item  LiveJournal is a blog-sharing website where users can maintain friend lists, keep a blog, journal or diary. Our data contains the friend lists for all users and their blog posts published from February 14th, 2010 to November 21st, 2011, which involves 9,573,127 users and 3,462,504 records of blog reference. Similar to Facebook, we depend on the friend list to build the underlying network topology. More importantly, LiveJournal users usually add URL links pointing to other relevant blogs when they refer them. As a result, we could use the URL reference to trace the information diffusion among users.

\end{itemize}

The originally constructed network is indicated by $\bar{G} =
\{\bar{V},\bar{E}\}$ in which $\bar{V}$ stands for the set of nodes
and $\bar{E}$ the set of edges. In the raw datasets of online social
platforms including Facebook and LiveJournal, we find many inactive
users who neither spread nor receive messages in network. Actually,
they just register an account but do nothing during the period of time
we collect data. Considering that no contributions are made by those
inactive nodes to the information diffusion process, we exclude them
from the original networks $\bar{G}$ and construct an active network
$\bar{G}_A = \{\bar{V}_A,\bar{E}_A\}$. Different from the online
social platforms, APS has no such inactive nodes as all the authors
have to publish papers and cite others' work. However, APS data
contains a minority of articles ($\sim 0.67\%$) whose number of
coauthors are so large (more than $100$ coauthors) that they would
produce extremely dense cliques. Therefore, we neglect those articles
in constructing the APS network. In all the social networks, we only
consider the largest connected component, denoted by $G =
\{V,E\}$. Properties of the original and truncated networks are
provided in Table \ref{tab:table1}.

\subsection*{Construction of Virtual Information Spreading}

In order to decide which strategy to use to locate the most influential nodes in networks, we intend to evaluate the collective influence exerted by the same number of influencers. The one that achieves the largest collective influence would be our first choice. To this end, the most straightforward idea is to compare the spreading range of multi-source spreading processes triggered by a fixed number of seeds selected by different methods. However, the multi-source spreading is an ideal process. In the ideal setting, multiple sources should be activated by the same piece of message at the same time. While in reality, such ideal situation rarely exists because of the intrinsic properties of real data. Users are interested in a wide range of topics, and they are receiving and delivering multifarious messages from time to time. It is unlikely that we can find enough real instances in which the spreaders happen to send out a same piece of message at the same time. Therefore, rather than enforcing real data to match the ideal expectation, we propose an alternative way - to construct a virtual multi-source spreading process.

\begin{figure}[ht]
\centering
\includegraphics[width=\linewidth]{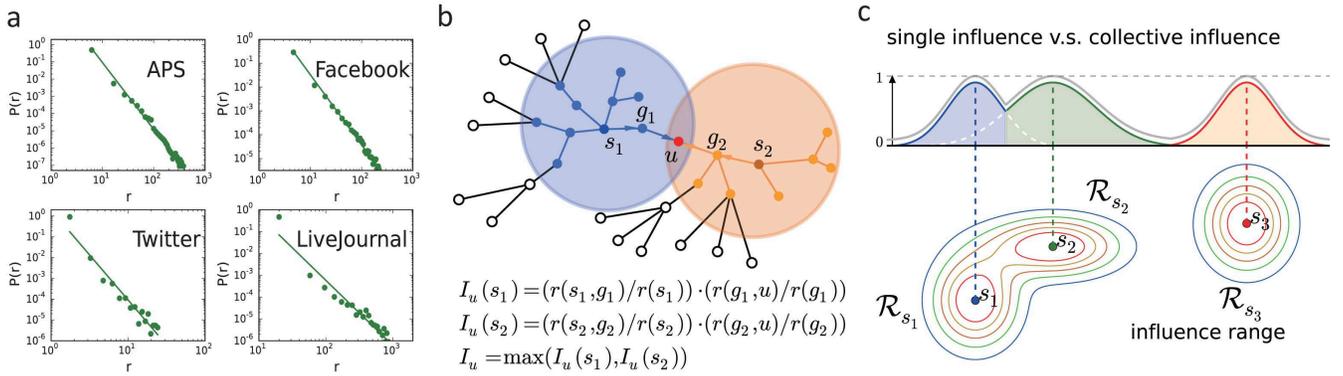}
\caption{Construction of virtual spreading based on people's interactions. \textbf{a,} Distribution of directed tie strength for real networks. The power law distribution demonstrates the heterogeneity of interactions between nodes. \textbf{b,} Calculation for influence strength. Nodes $s_1$ and $s_2$ are two distinct spreaders, the maximum spreading layer is set as $L = 2$. Node $u$ is influenced by two seeds with the strength $I_u(s_1)$ and $I_u(s_2)$. We select the largest value to indicate the collective influence enforcing on it. \textbf{c,} An illustration of single influence strength $I_u(s)$ along with collective influence strength $I_u$. The three circle-like areas represent the corresponding influence ranges $\mathit{R}_{s_1}, \mathit{R}_{s_2}, \mathit{R}_{s_3}$ for distinct spreaders $s_1, s_2, s_3$, and the contour lines indicate the levels of influence strength $I_u$. When projecting it onto $2$-dimensional space, we have the corresponding distribution. The collective outcome $I_u$ (indicated by gray curve) is obtained by combining the single influence strengths of all the spreaders.}
\label{fig:description}
\end{figure}

The main idea behind the virtual multi-source spreading processes is that users are expected to follow the behavioral patterns expressed in real data \cite{pei2015exploring}. For user $i$ with $k_i$ neighbors who have chances to access information from $i$, the closely-tied neighbors interested in user $i$'s publications or posts would be more likely to inherit messages from $i$. On the contrary, those weakly-tied friends would occasionally be influenced by the information released from $i$. To reflect this effect, we propose a notion named the strength of directed ties $r$. For a directed link from $i$ to $j$, the strength $r(i,j)$ is defined by the number of messages, e.g., publications or posts, passed from $i$ to $j$. By definition, the strength of directed tie $r(i,j)$ from $i$ to $j$ is not generally equal to $r(j,i)$ from $j$ to $i$. Figure \ref{fig:description}a reveals that the strength of directed tie follows a power-law distribution. We assume that, in the virtual processes, people would continue to maintain such behavioral patterns. In this way, we can approximate the multi-source information diffusion and obtain the collective influence as follows.

In the virtual processes, suppose a $q$-percentage of initial
spreaders are activated at the beginning, denoted by $S = \{s_i \mid i
= 1,2,...,n, n = N\cdot q\}$. We introduce a quantity $I_u(s) \in
\left[0, 1\right]$ to represent the single influence strength that
node $u$ is affected by spreader $s$. Correspondingly, we employ $I_u$
to indicate the collective influence strength enforced by all seeds
$S$. Both of their calculations can rely on the above mentioned
strength of directed ties (shown in Figure
\ref{fig:description}b). For an arbitrary spreader $s$, the influence
strength $I_{g_1}(s)$ from $s$ to its neighbor $g_1$ depends on the
strength of directed tie $r(s,g_1)$, or in other words, depends on the
tendency of $g_1$ to receive information from $s$. Assume that, during
one period of time, $s$ has totally sent out $r(s)$ pieces of messages
and $g_1$ has accepted $r(s,g_1)$ of them [$r(s,g_1) \leq r(s)$]. The
proportion of acceptance $r(s,g_1)/r(s)$ can be viewed as a proxy of
influence strength from $s$ to $g_1$, i.e. $I_{g_1}(s) = r(s,g_1)/r(s)$. 
Next, $g_1$ might affect its neighbor $g_2 \ne s$ in
the same way. Then we follow the spreading paths, multiply the
proportions together and then acquire the influence strength $s$
enforcing on its $l$-step neighbor $g_l$, say
\begin{equation}
\label{eq:1}
I_{g_l}(s) = \prod_{k=1}^{l}r(g_{k-1},g_k)/r(g_{k-1}),
\end{equation}
where $g_0 = s$. Figure \ref{fig:description}b gives an example with $l = 2$. As none of messages can spread
infinitely, we set a number $L$ as the maximum layer of spreading, so
that the influence range, denoted by $\mathit{R}_s$, could be
approximated by a ball around $s$ with the radius $L$ (shown in Figure
\ref{fig:description}c). Within each $\mathit{R}_s$, we have
$I_{g_0}(s) = 1$ for the central spreader $s$, then the value
decreases as $l$ becoming larger, and $I_{g_l}(s) = 0 \left(l >
L\right)$ for any external node. The schematic diagram regarding the
distribution of influence strength within $\mathit{R}_s$ can be seen
in Figure \ref{fig:description}c. For APS and LiveJournal data, we
know more information about references, the detailed calculation of
influence strength is shown in \textbf{Methods}.

To obtain the collective influence $I_u$ for node $u$, we apply
\begin{equation}
\label{eq:2}
I_u = {\max}_{i=1}^{n}{I_u(s_i)}.
\end{equation}
Referring to Figure \ref{fig:description}b,c, it is straightforward to understand when node $u$ does not belong to any influence range, $I_u(s_i) = 0$ for any $i$, in which case the collective influence should be zero. For the case that node $u$ is only influenced by one spreader, for example $I_u(s_i) > 0$ and $I_u(s_j) = 0$ for any $j \ne i$, the collective influence should be chosen as the positive (largest) one $I_u = I_u(s_i)$. More generally, if node $u$ lies within the overlapping areas of more than one influence ranges, i.e. it is affected by more than one sources, we ought to choose the largest potential influence to be its collective influence during the virtual spreading process. Finally, we sum up all the $\{I_u \mid u = 1,2,...,N\}$ together to obtain the collective influence that spreaders impose on the entire system through
\begin{equation}
\label{eq:3}
Q(q) = \sum_{u=1}^{N}I_u/N.
\end{equation}
Since $0 \leq I_u \leq 1$, we have $0 \leq Q(q) \leq 1$, which corresponds to the collective spreading range for the virtual process (see Figure \ref{fig:description}c).

In general, the virtual process of multi-source spreading constructed
here is an approximation of real information diffusion. We take
advantage of real data to extract users' behavioral patterns, base on
which, we can calculate the single influence and collective influence
that spreaders impose on each node. Given that, we can finally compute
the collective influence exerted by all influencers on the entire
network.

\subsection*{Comparison of Different Methods}

In this section, we compare CI algorithm with four other widely-used
heuristic measures, including adaptive high-degree (HDA)
\cite{morone2015influence}, high-degree(HD)
\cite{albert2000error,pastor2001epidemic}, PageRank (PR)
\cite{brin1998the} and k-core
\cite{dorogovtsev2006kcore,carmi2007amodel,kitsak2010identification,pei2014searching}
(details about methods are shown in \textbf{Methods}). Recall that,
our first step is to identify the $q$-percentage of initial spreaders
according to different methods. Secondly, we construct a virtual
multi-source spreading process. Finally, we compare the virtual
spreading range $Q(q)$, i.e. the collective influence of those initial
influencers.

\begin{figure}[ht]
\centering
\includegraphics[width=0.7\columnwidth]{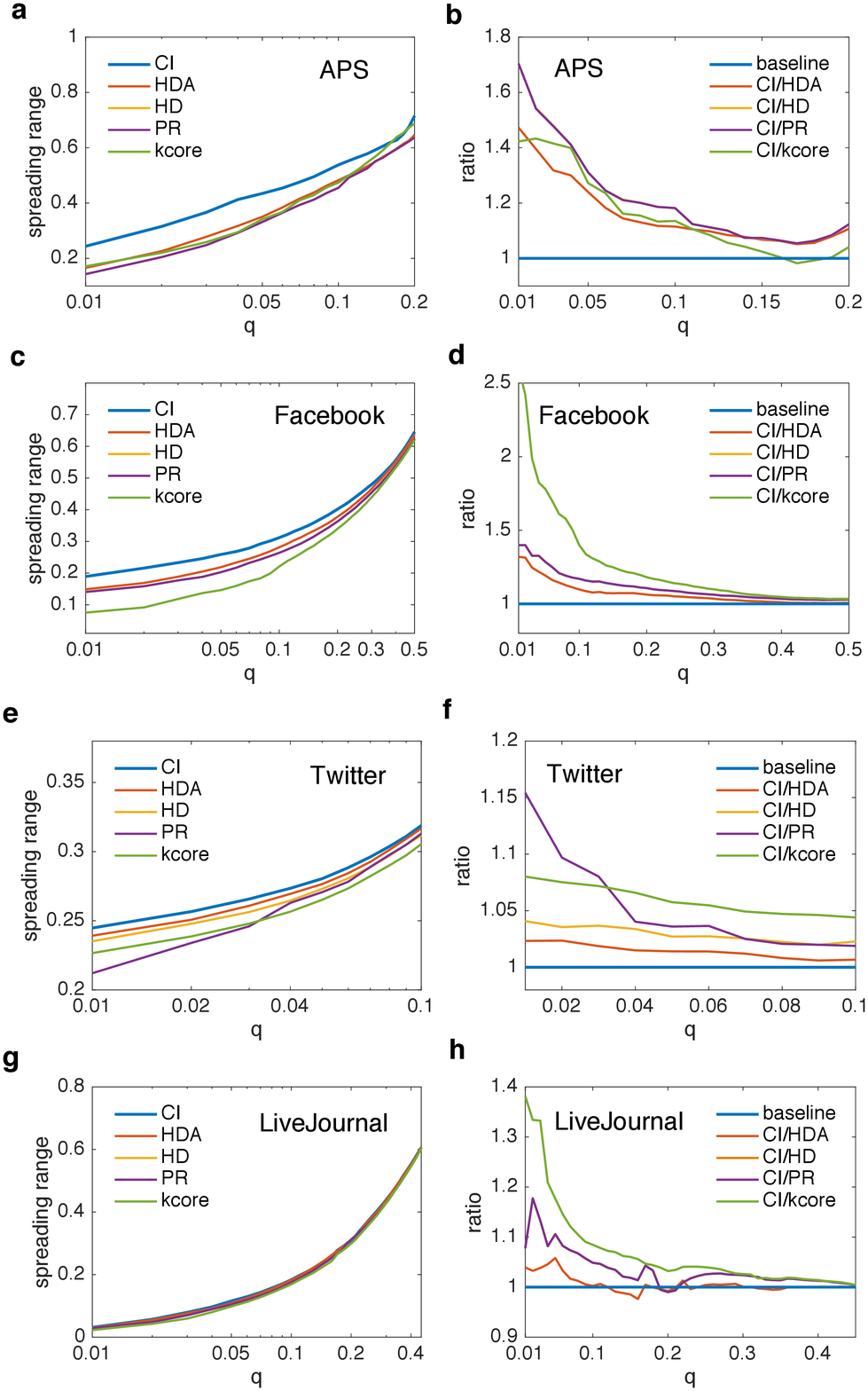}
\caption{Performance of CI in large-scale real social networks. The datasets contain APS (\textbf{a,b}), Facebook (\textbf{c,d}), Twitter (\textbf{e,f}) and LiveJournal (\textbf{g,h}). We compare the virtual spreading ranges of different methods in \textbf{a,c,e,f}. With a fixed fraction $q$ of seeds, CI's virtual spreading range is larger than all the heuristic approaches. Besides, we also show the ratios of spreading ranges between CI and others in \textbf{b,d,f,h}. It reveals that the ratios are always larger than $1$ (higher than the baseline), implying that CI is an effective strategy in locating multiple spreaders. We set $L = 3$ for APS and Facebook which have large value of $\langle k^d \rangle$, and $L = 5$ for Twitter and LiveJournal that have small value of $\langle k^d \rangle$. We care about the results when $q$ is small, so we limit $q$ within the range of small value. As $q$ increases, the performances of all the strategies become similar.}
\label{fig:sp_r}
\end{figure}

Figure \ref{fig:sp_r}a,c,e,g show the virtual collective influence scores obtained by CI, HDA, HD, PR and k-core for the four networks -- APS, Facebook, Twitter and LiveJournal. It can be seen that for a certain value of $q$, the set of nodes selected by CI can diffuse the information to a larger scale of populations than those obtained by other methods. CI's good performance is more prominent for APS and Facebook data as their diffusion instances are relatively abundant. To clearly distinguish the performances of different methods, we also present the ratios between CI's collective influence score and those of other approaches (Figure \ref{fig:sp_r}b,d,f,h). It reveals that the ratios are always larger than one (indicated by the baseline at $1$) for all datasets. Besides, the ratio is relatively large when $q$ is small. As $q$ increases, it would decline accordingly, suggesting that if we select a larger amount of influencers, the collective influence score obtained by all methods would become similar. Among the competing heuristic methods, HDA can be viewed as a special case of CI with the calculation radius being zero \cite{morone2015influence} (see \textbf{Methods}). However, HDA's capability in locating influencers is limited by the lack of knowledge of the surrounding nodes, so it is a strategy obtained from the non-interacting point of view. K-core method, a good predicator for locating single ``superspreaders'' \cite{kitsak2010identification,pei2014searching}, whereas fails to identify multiple spreaders in the multi-source spreading process. This is because the selected influential nodes tend to cluster together in the core shells which induces large overlapping of their influence areas.

\begin{figure}[ht]
\centering
\includegraphics[width=0.8\columnwidth]{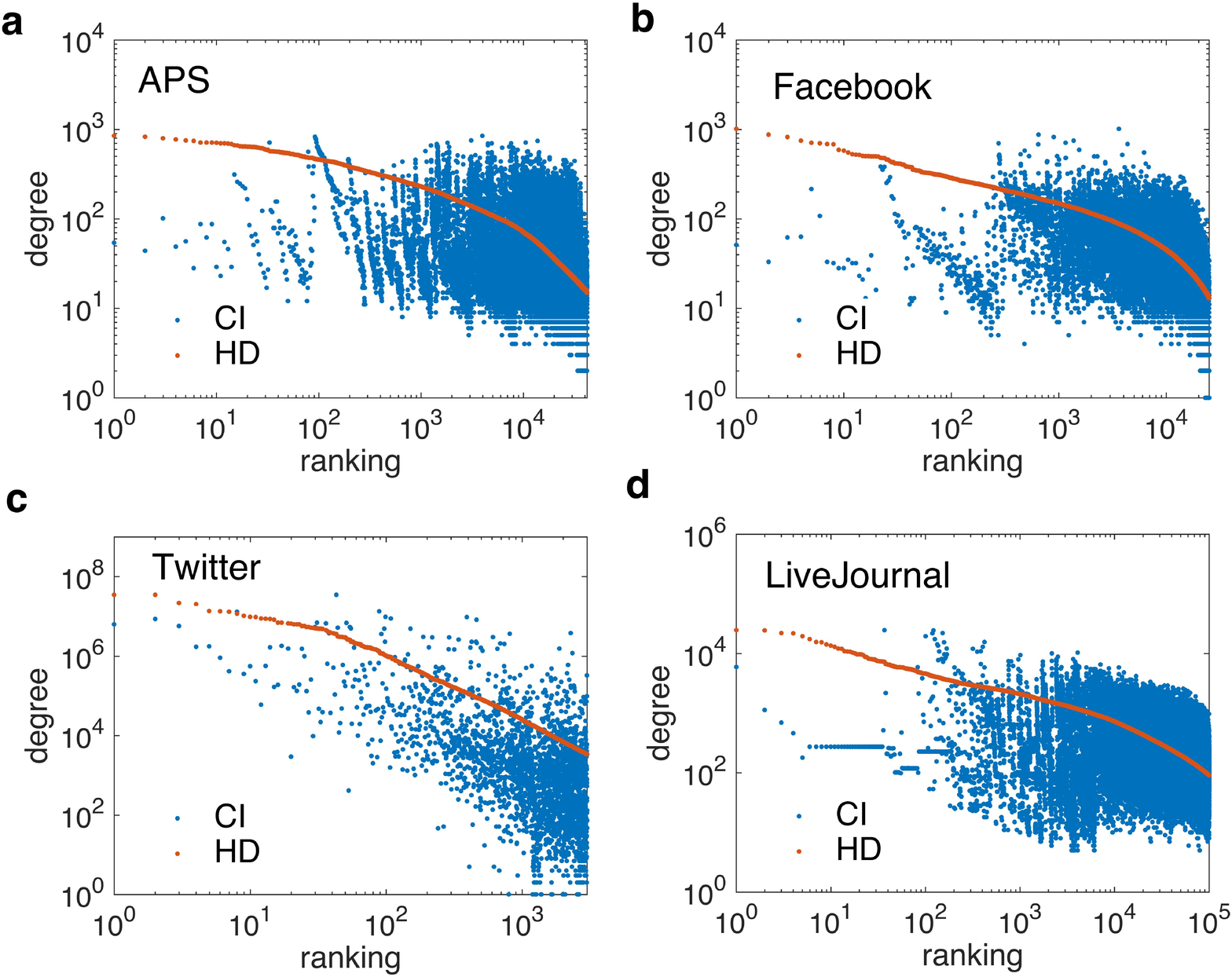}
\caption{Degree versus ranking. We show the degrees of nodes ranked
  (from highest to lowest) by CI and HD for APS (\textbf{a}), Facebook
  (\textbf{b}), Twitter (\textbf{c}) and LiveJournal (\textbf{d}). It
  shows that CI can find those previously neglected weak nodes to
  emerge among most significant influencers. Meanwhile, some most
  connected nodes are ranked as moderate influencers by CI,
  indicating that such weak node effect is a consequence of collective
  influence in the case of multiple spreaders. This result has
  important consequences for ranking of researchers in scientific
  networks.}
\label{fig:deg}
\end{figure}

\begin{figure}[ht]
\centering
\includegraphics[width=0.8\columnwidth]{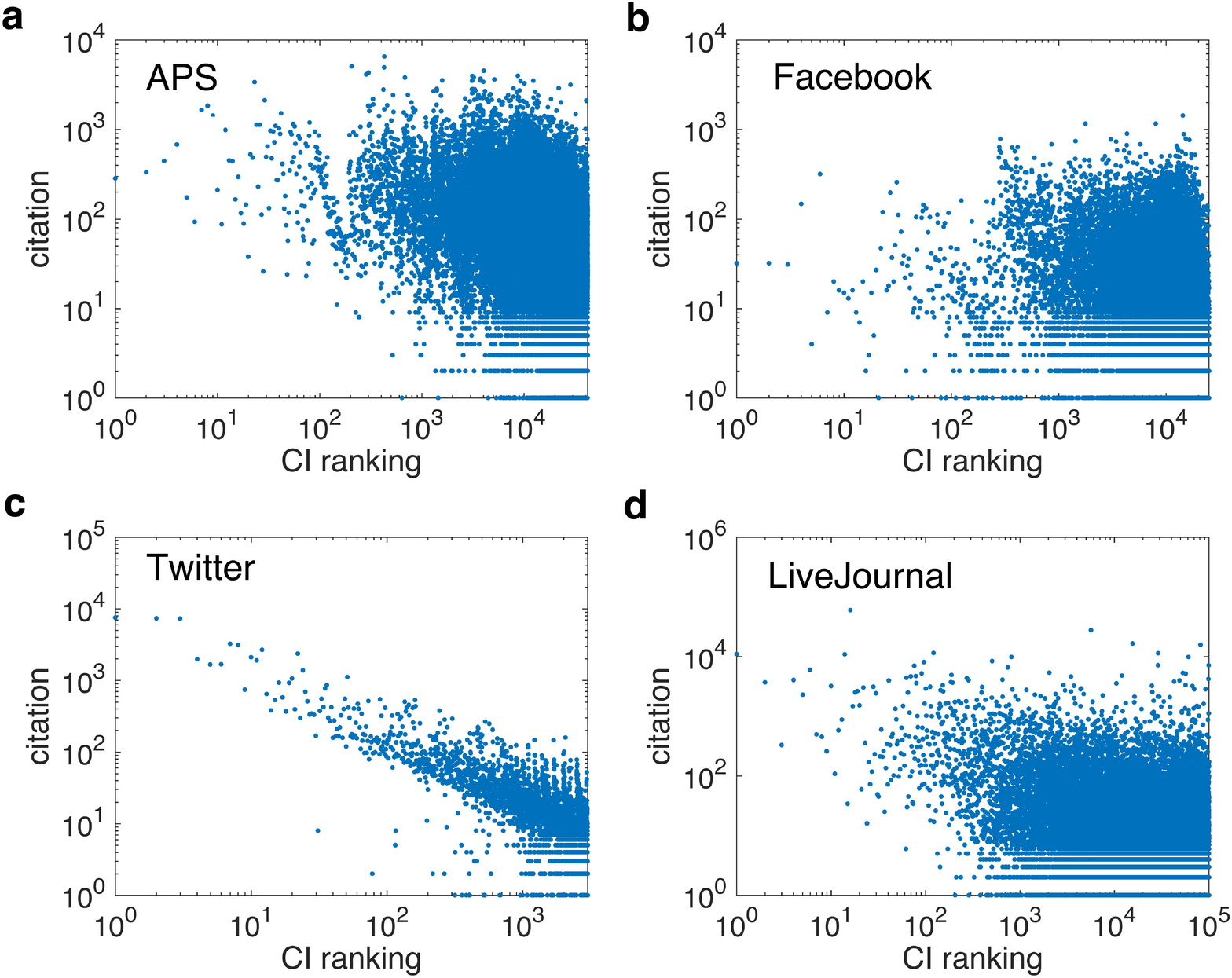}
\caption{The number of citations versus CI ranking. We present the
  number of citations (comments, reposts or references) of nodes
  ranked by CI strategy for APS (\textbf{a}), Facebook (\textbf{b}),
  Twitter (\textbf{c}) and LiveJournal (\textbf{d}). Despite that in
  Twitter data, the most influential user is exactly the one with the
  largest amount of citations, the overall results still prove that
  large number of citations is not necessarily a reliable measure for
  identification of top-ranking influencers. This fact has meaning especially for academic rankings for physicists in community like APS. CI takes into account the maximization of influence in the
  whole network of each scientist rather than just the local
  information given by the number of citations. Thus a highly cited
  author may not have a large impact in the community if he/she is
  isolated in the periphery. An optimal measure as CI should rank such
  a scientist lower in the scientific community. This result calls for
  a revision of rankings based solely on the local information rather
  than the collective influence in the entire network community. We
  elaborate more on this problem in subsequent publications.}
\label{fig:citation}
\end{figure}

Besides, we also investigate the characteristics of influencers that
CI has identified. Figure \ref{fig:deg}a shows the degree comparison
of nodes ranked by CI and HD (from the most influential to the
least). Unlike HD finding influencers just relying on degree, CI's
most important nodes contain not only hubs but also many
weakly-connected nodes. Besides, some of the most connected nodes turn
out to be moderate influencers. It confirms the former conclusion that
collective influence is determined by the interplay of all the
influencers. Under certain circumstances, some low-degree nodes
surrounded by hierarchical coronas of hubs have larger contributions
to collective influence than those high-degree nodes connecting to
peripheral leaves \cite{morone2015influence}. In addition, we have
also examined the correlation between CI ranking and the number of
citations in Figure \ref{fig:citation}. The number of citation for
each user is defined as how many times other people have accepted or
inherited information from him/her directly. We acquire such
information through checking the citations (APS), comments (Facebook),
retweets (Twitter) as well as URLs reference (LiveJournal). Except for
Twitter, the other datasets show us that the most influential
nodes are not necessarily those with the largest number of
citations. The uniqueness of Twitter might be explained by considering the mechanism of network formation and the way of data collection. Twitter platform facilitates users arbitrarily following others, making it possible that super hubs with millions of followers emerge and hold significant influence; Besides, Twitter is gathered by focusing on a popular topic "Donald Trump", the topic-based data might easily detect those extremely popular users who also play important role in spreading. Therefore, the phenomenon shown in APS, Facebook and LiveJournal suggest practical implications for academic rankings. When evaluating a researcher's scientific impact within a field, his/her number of citation is not the determinative factor
\cite{wang2013quantifying,radicchi2009diffusion}. It also reminds us
that influence is an emergent property arising from interactions
rather than an evaluation by viewing nodes individually.

\section*{Discussion}

It is of importance to search for the most influential nodes in
social networks. For a long time, heuristic approaches have been
widely used to find superspreaders, yet without an ultimate solution
for finding multiple influencers. Recently, a
rigorous framework called collective influence (CI), along with a scalable algorithm, has been put forward to resolve the many-body problem. Even though CI has been shown to be
effective in percolation model, we still need to verify its
performance particularly in the real case of information diffusion. To
achieve this, we collect data from four social media -- APS journals,
Facebook, Twitter as well as LiveJournal platforms. Different from the
situation of finding single superspreaders where we check each node's spreading range, under the circumstance of multiple spreaders, we should examine the collective spreading range. Given the
difficulty that ideal multi-source spreading processes triggered by
same messages at the same time are scarce in real-world diffusion, we
propose a virtual multi-source spreading according to users'
behavioral patterns to approximate the ideal process. Finally, by
comparing the collective influence, i.e. the spreading ranges in
virtual process, we find that CI is effective in finding
multiple influencers.

Moreover, our finding indicates that quantities from a non-interacting
viewpoint, such as degree and the number of citations, are not
reliable in measuring nodes' importance in collective influence. Our
investigation for influencers' properties confirms that
influence is an effect of cooperation in multi-source spreading. Our
results can be transformed into an effective way to rank scientist
in academic communities according to their collective influence rather
than on the commonly used local connectivity metric, like the number
of citations or collaborations in the H-index (Hirsch number). Using
the number of citations, as shown in Fig. \ref{fig:citation}, can be
a poor indicator of the collective influence of a researcher on other
researchers in the community. A global quantity like the
Collective Influence that takes into account the optimization of
influence of all researchers at once, provides a meaningful
ranking of researchers according to the maximization of their
influence. More studies will follow to elaborate on this particular
point.

\section*{Methods}

\subsection*{Collective Influence Method}

\textbf{Collective Influence (CI) Algorithm} \cite{morone2015influence}. CI is an optimization algorithm that aims to find the minimal set of nodes that could fragment the network in optimal percolation. In percolation theory \cite{bollobas2006percolation}, if we remove nodes randomly, the network would undergo a structural collapse at a critical fraction where the probability that the giant connected component exists is $G = 0$. The optimal percolation is an optimization problem which attempts to find the minimal fraction of influencers $q_c$ to achieve the result $G(q_c) = 0$.
Let the vector $\textbf{n} = (n_1, n_2, ..., n_N)$ represent whether a node is removed ($n_i = 0$) or not ($n_i = 1$), and the vector $\textbf{v} = (v_1, v_2, ..., v_N)$ represent whether a node belongs to the giant connected component ($v_i = 1$) or not ($v_i = 0$). The relationship between $\textbf{n}$ and $\textbf{v}$ can be derived in locally tree-like networks using message passing (MP) approach \cite{bianconi2014multiple,karrer2014percolation}:
\begin{equation}
\label{eq:4}
v_{i \to j} = n_i[1-\prod_{k \in \partial i\backslash j}(1-v_{k \to i})],
\end{equation}
where $v_{i \to j}$ indicates the probability of $i$ being in the giant component when $j$ is absent, and $\partial i\backslash j$ is the neighbors of $i$ besides $j$. The equation's possible solution $v_{i\to j} = 0$ for all $i \to j$ is associated with the special situation where the giant connected component is absent; therefore, to obtain $G(q) = 0$, the stability of this solution must be guaranteed. As a matter of fact, the stability of $v_{i \to j} = 0$ is controlled by the largest eigenvalue $\lambda (\textbf{n};q)$ of the linear operator $\hat{\mathit{M}}$, which is defined on the directed edges of networks as
\begin{equation}
\label{eq:5}
\mathit{M}_{k \to l, i \to j} \equiv \frac{\partial v_{i \to j}}{\partial v_{k \to l}}|_{\{v_{i \to j}=0\}}.
\end{equation}
It can be expressed as
\begin{equation}
\label{eq:6}
\mathit{M}_{k \to l, i \to j} = n_i \mathit{B}_{k \to l, i \to j},
\end{equation}
where $\mathit{B}_{k \to l, i \to j}$ is the non-backtracking matrix of the network \cite{hashimoto1989zeta,angel2015the}. $\mathit{B}$ stores the topological interconnections of network whose element $\mathit{B}_{k \to l, i \to j} = 1$ if $l = i, j \ne k$. So far, the original optimal percolation problem has been rephrased as a mathematical statement: finding the optimal configuration of $\textbf{n}^*$ with size $q_c$ that achieves the critical threshold:
\begin{equation}
\label{eq:7}
\lambda(\textbf{n}^*;q_c) = 1.
\end{equation}
The eigenvalue $\lambda (\textbf{n};q)$ can be calculated according to power method \cite{bhatia2002stability}:
\begin{equation}
\label{eq:8}
\lambda(\textbf{n}) = \displaystyle\lim_{l \to \infty}\left[\frac{|\textbf{w}_l(\textbf{n})|}{|\textbf{w}_0|}\right]^{1/l}.
\end{equation}
At a finite $l$, $|\textbf{w}_l(\textbf{n})|^2$ is the cost energy function of influence that needs to be minimized. Take Equation \ref{eq:8} as a starting point, the problem of finding the optimal set of influencers can be solved by minimizing the following cost function:
\begin{equation}
\label{eq:9}
E_l\left(\textbf{n}\right) = \displaystyle\sum_{i=1}^N \left(k_i-1\right) \sum_{j\in \partial \mathrm{Ball}(i,l)} \left(\prod_{k\in\mathit{P}_l(i,j)}n_k\right)(k_j-1),
\end{equation}
where $\mathrm{Ball}(i,l)$ is the set of nodes inside the ball of radius $i$ around the central node $i$, and $\mathit{P}_l(i,j)$ is the shortest path of length $l$ connecting $i$ and $j$. To minimize the energy function of a many-body system, an adaptive method is developed with the main idea of removing the nodes causing the biggest drop in the energy function - CI algorithm.
In general, CI algorithm can be stated as follows. Firstly, it
considers the nodes at the frontier $j \in \partial
\mathrm{Ball}(i,l)$ and assigns to node $i$ a collective influence
value at the level of $l$ as
\begin{equation}
\label{eq:10}
\mathrm{CI}_l(i) = (k_i-1)\sum_{j\in\partial \mathrm{Ball}(i,l)} (k_j-1).
\end{equation}
Starting with the node with the highest $\mathrm{CI}_l$, CI adaptively removes nodes and after
each removal, it recalculates $\mathrm{CI}_l$ for all the rest nodes
in the system. From the calculation we know that CI has richer
topological contents and its performance will be improved as $l$
increases, but no larger than the network diameter because this case
amounts to random identification. In our analysis, we adopt the
parameter $L = 3$ in the adaptive calculation of CI, which has been
shown to be sufficient for optimal percolation. At the opposite
extreme $l = 0$, we have $\mathrm{CI}_{0}(i) = (k_i-1)^2$. Under this
situation, CI algorithm is reduced to the \textbf{High-degree adaptive
  (HDA) method}. For $l \geq 1$, CI also considers the surrounding neighborhoods and the
interactions among nodes; meanwhile, it is an easily-implemented
algorithm as it only needs local topological structure within the ball
of the radius $l$ instead of the whole network structure. More
importantly, its computational complexity is $\mathnormal{O}(N\log N)$, which guarantees its application for large
real networks \cite{flaviano}.

\subsection*{Heuristic Methods}

\textbf{k-core} \cite{dorogovtsev2006kcore,carmi2007amodel,kitsak2010identification,pei2014searching}. In k-core method, nodes are ranked based on their $k_S$ values, which are calculated during the process of $k$-shell decomposition. In $k$-shell decomposition, nodes are removed iteratively. Firstly, nodes with $k = 1$ are removed and continue pruning the networks until no leaf nodes are available. The set of removed nodes compose the peripheral k-shell with index $k_S = 1$. Similarly, the next k-shells with index $k_S>1$ are generated and the nodes located within the core area have the highest $k_S$ values. Actually, in $k$-shell composition, all the nodes are divided into different shells according to their relative locations in networks. Compared with the peripheral nodes, the core nodes have higher probabilities to cause large-scale diffusions. This method has been revealed to perform well in searching for single spreaders who can yield large influence areas. However, it has a poor performance when being used to optimizing the collective spreading caused by multiple spreaders \cite{kitsak2010identification}. Because k-core would select a bunch of nodes within or near the network core, so their influence areas would heavily overlap and produce a bad collective outcome \cite{kitsak2010identification}.

\textbf{PageRank(PR)} \cite{brin1998the}. PageRank algorithm was firstly proposed by S. Brin and L. Page and used by Google in order to rank websites. It extends the idea in academic citation that the number of citations or backlinks give some approximation of a page's importance, by not counting links equally but normalizing by the number of links on a page. Its calculation is as follows: if page $A$ has pages $T_1, ..., T_N$ citations with the associated PageRank as $\mathrm{PR}(T_1), ..., \mathrm{PR}(T_N)$, then the PageRank of $A$ is given by
\begin{equation}
\label{eq:11}
\mathrm{PR}(A) = (1-d) + d \left(\frac{\mathrm{PR}(T_1)}{C(T_1)} + ... + \frac{\mathrm{PR}(T_N)}{C(T_N)}\right),
\end{equation}
in which $C(A)$ is defined as the number of links going out of page $A$. PageRank outputs a probability distribution used to represent the likelihood that a person randomly clicking on links will arrive at any particular page. The higher the probability, the higher the PR value of this page. In practice, PageRank can be calculated using a simple iterative algorithm and corresponding to the principal eigenvector of the normalized link matrix of the web network.

\textbf{High-Degree(HD)} \cite{albert2000error,pastor2001epidemic}. HD method ranks nodes directly according to the number of connections. Compared with other methods requiring global network structures like k-core and PageRank, HD only needs local information and is easily implemented. However, it cannot deal with the circumstance in which hubs form tight community such that their spreading areas would heavily overlap \cite{wasserman1944social,colizza2006detecting}.

\textbf{High-Degree Adaptive(HDA)}. HDA is the refined adaptive
version of HD method. To help mitigate the above mentioned situation,
HDA recalculates the degrees after each removal. It can also be viewed
as a special case of CI algorithm at $l = 0$. Compared with CI, HDA
represents the one-body scenario where the influencers are considered
in isolation and therefore, it lacks the collective influence effects
from the neighborhood.

\subsection*{Data Processing}

\textbf{Analyzing APS and LiveJournal}. In terms of APS, we know the specific article pairs $(\alpha,\beta)$, which means paper $\alpha$ cites paper $\beta$, In other word, the authors $A_{\beta}$ of $\beta$ spread their scientific discoveries to the authors $A_{\alpha}$ of $\alpha$. Therefore, for an arbitrary author $s$, we can know his or her journal set $J(s) = {\{J_i\mid i = 1,2,...,n_s\}}$ in which $J_i$ indicates each piece of paper and $n_s$ stands for the number of papers published by $s$. By tracking the spreading for each paper $J_i$ through citation flows, we can determine its influence range $\mathit{R}_{s}(J_i)$ containing all people who have cited this paper $J_i$. For each receiver $u \in \mathit{R}_s = \{\mathit{R}_s(J_i)\mid i = 1,2,...,n_s\}$, we calculate the individual influence strength by $I_u(s) = (\sum_{i=1}^{n_s}\delta_{u\in \mathit{R}_{s}(J_i)})/n_s$ where $\delta_{u\in \mathit{R}_{s}(J_i)} = 1$ if and only if $u\in \mathit{R}_{s}(J_i)$. Large values of $I_u(s)$ means that $u$ is more likely to cite the work of $s$ than other peers. Next, the collective influence strength from all sources can be obtained by $I_u = \it{\max_{i=1}^{n}}{I_u(s_i)}$. In LiveJournal, we know information about blog references. So, we can follow the similar method as in APS to process LiveJournal data.

\section*{Acknowledgements}

This work was supported by NIH-NIGMS 1R21GM107641, NSF-PoLS PHY-1305476 and ARL Cooperative Agreement Number W911NF-09-2-0053, the ARL Network Science CTA. We thank Lev Muchnik for providing the data on LiveJournal.

\section*{Author contributions statement}

H.A.M. designed research; X.T., S.P. , F.M., H.A.M. analyzed data, prepared figures and wrote the main manuscript text; All authors reviewed the manuscript.

\section*{Additional information}
\subsection*{Competing financial interests}
The authors declare no competing financial interests.

\begin{table}[ht]
\centering
\begin{tabular}{c|ccccccccc}
Networks & $\bar{N}$ & $\bar{M}$ & $\bar{N}_A$ & $\bar{M}_A$ & $N$ & $M$ & $\langle k \rangle$ & $\langle k^d \rangle$ & $q_c$ \\
\hline
\hline
APS & 230,521 & 1,607,305 & 230,521 & 1,607,305 & 190,161 & 1,582,710 & 16.4 & 37.4 & 20\% \\
Facebook & 63,731 & 817,090 & 45,746 & 703,924 & 45,459 & 703,803 & 31.0 & 18.8 & 45\% \\
Twitter & 311,334 & 151,654 & 311,334 & 151,654 & 29,463 & 143,220 & 9.7 & 5.1 & 6\% \\
LiveJournal & 9,573,126 &188,240,039 & 304,858 & 19,785,460 & 290,362 & 19,783,730 & 136.3 & 7.7 & 46\% \\
\hline
\end{tabular}
\caption{\label{tab:table1}Properties of the original and processed
  networks $\bar{G}, \bar{G}_A, G$ in this article. In the table,
  $\bar{N}$ ($\bar{M}$) is the number of nodes (edges) in the original
  networks $\bar{G}$, $\bar{N}_A$ ($\bar{M}_A$) represents the number
  of nodes (edges) in the active network $\bar{G}_A$, $N$ ($M$)
  indicates the number of nodes (edges) in the network $G$. $\langle k
  \rangle$ is the average degree of network $G$. $\langle k^d \rangle$
  denotes the average out-degree of diffusion graph, i.e. the average
  number of messages which have been sent out. Besides, $q_c$
  indicates CI's minimal fraction of influencers to fragment the
  networks in optimal percolation \cite{morone2015influence}. All
  datasets are available at \url{kcore-analytics.com}.}
\end{table}

\end{document}